\documentclass[prl,twocolumn,superscriptaddress]{revtex4-2}
\usepackage[utf8]{inputenc}
\usepackage{amsmath,amssymb}
\usepackage{graphicx}
\usepackage{xcolor}
\usepackage{verbatim}

\begin{document}

\title{Laboratory constraints on ultralight axion-like particles from precision atomic spectroscopy}
\author{Joshua Berger}
\affiliation{Colorado State University, Fort Collins, Colorado 80523, USA}
\author{Amit Bhoonah}
\affiliation{Department of Physics and Astronomy,  University of Pittsburgh, Pittsburgh, PA 15260, USA}
\begin{abstract}
    Ultralight bosonic dark matter has come under increasing scrutiny as a dark matter candidate that has the potential to resolve puzzles in astronomical observation. We demonstrate that high-precision measurements of time variation in the frequency ratios of atomic transitions achieves leading sensitivity to ultralight axion-like particle dark matter at low masses. These bounds are the first laboratory-based bounds on this class of dark matter models. We propose further measurements that could enhance sensitivity to ultralight axion-like particles.
\end{abstract}
\maketitle

While abundant evidence for the existence of dark matter has been accumulated over the last several decades, its particle nature and non-gravitational interactions remain elusive. Many models have been developed to resolve this mystery, but no definitive laboratory discovery has been made. Constraints on weakly interacting massive particles (WIMPs), historically the most popular dark matter candidate, continue to improve, but it remains important to find ways of searching for other dark matter candidates. 

Axion Like Particles (ALPs) are one such WIMP alternative. Historically, they emerged from efforts to explain the lack of CP violation in the QCD sector \cite{Peccei:1977hh}, the so-called \textit{strong CP problem}. The Peccei-Quinn (PQ) mechanism, as it became known, works by introducing a new spontaneously broken global symmetry which relaxes the value of the (CP violating) $\bar{\theta}$ parameter of QCD to zero. Since instanton effects break this \textit{PQ symmetry} explicitly, there is an associated light pseudo-Goldstone boson, the \textit{axion}. Early realisations with PQ symmetry breaking close to the weak scale \cite{Wilczek:1977pj,Weinberg:1977ma} were quickly ruled out, but later models of the ``invisible axion" \cite{Kim:1979if,Shifman:1979if,Dine:1981rt,Zhitnitsky:1980tq} remain viable solutions to the strong CP problem. The latter are also defined by their feeble interactions with normal matter such as electrons, nucleons, and especially photons, making them plausible dark matter candidates \cite{Preskill:1982cy,Abbott:1982af,Dine:1982ah}.  

Requiring that axions both solve the strong CP problem and make up the dark matter in the universe places severe constraints on the ALP parameter space. For clarity, we will refer to these as \textit{QCD axion} dark matter. One may relax the strong CP assumption and only require that ALPs be dark matter, which drastically opens up the parameter space, particularly the ultralight window, which is not possible for QCD axions. Motivation for ultralight dark matter (ULDM) comes from astronomical observations of galactic-scale structure. Puzzles such as the core-cusp \cite{1994Natur.370..629M}, too-big-to-fail \cite{2011MNRAS.415L..40B}, and missing satellite \cite{1993MNRAS.264..201K,1999ApJ...522...82K} problems have motivated thinking about dark matter as an ultralight boson with a Compton wavelength comparable to the small scale structures in question, which corresponds to a mass of order $10^{-21}~\text{eV}/c^2$. These puzzles are not without controversy, but modelling the dark matter as an ultralight boson modifies the small scale structure of dark matter and alleviates them. 

Compared to WIMPs, ULDM has drastically different laboratory search prospects. At such low masses the energy transfer from ultralight bosons to matter falls well below even the most optimistic detection thresholds for traditional dark matter experiments. Hence, with very few exceptions, constraints on ULDM have been either astrophysical or cosmological. Since the latter are not free from assumptions about cosmological history or astrophysical modelling, direct laboratory searches for ULDM remain invaluable but challenging as they require the development of novel techniques that differ drastically from those used at most direct detection experiments, where dark matter is assumed to behave like a collection of individual particles. For ULDM, on the other hand, the number densities in question are so large that a wave-like behaviour is more appropriate, which laboratory searches must exploit.

In this Letter, we propose using high precision atomic spectroscopy to perform a laboratory search for ultralight ALPs. Our work establishes the first laboratory constraints on the latter for some of the parameter space that had so far only been explored using cosmological or astrophysical systems. We also explore how further techniques from high precision atomic spectroscopy can be used to search for ultralight ALPs and establish a new connection between the fields of elementary particle and atomic, molecular, optical physics.

{\bf Model $\&$ Phenomenology.} The most interesting coupling for ALPs from a phenomenological perspective is their electromagnetic interaction, 
\begin{equation}
		\mathcal{L}_{a\gamma\gamma} =  - \frac{1}{4 \mu_0} g_{a\gamma\gamma} a F_{\mu\nu} \widetilde{F}^{\mu\nu} = \sqrt{\frac{\epsilon_0}{\mu_0}} g_{a\gamma\gamma} a  \mathbf{E} \cdot \mathbf{B},
	\end{equation}
where $a$ represents the axion field of mass $m_{a}$, $F_{\mu\nu}$ the photon kinetic operator, $\widetilde{F}_{\mu\nu} = (1/2) \epsilon_{\mu\nu\rho\sigma} F^{\rho\sigma}$, and $g_{a\gamma\gamma}$ the dimensionful axion-photon coupling constant. The corresponding five equations of motion are known as the equations of axion electrodynamics \cite{PhysRevLett.58.1799},
\begin{eqnarray}
		\nabla \cdot \mathbf{E} & = & \frac{\rho}{\epsilon_0} - c g_{a\gamma\gamma} \nabla a \cdot \mathbf{B} \label{eq:max1} \\
		\nabla \cdot \mathbf{B}& = & 0 \label{eq:max2} \\
		\nabla \times \mathbf{B}- \frac{1}{c^2} \frac{\partial \mathbf{E}}{\partial t} & = & \mu_0 \mathbf{J} +  \frac{g_{a\gamma\gamma}}{c}\left( \frac{\partial a}{\partial t} \mathbf{B}+ \nabla a \times \mathbf{E}\right)\label{eq:max3} \\
		\nabla \times \mathbf{E}+ \frac{\partial \mathbf{B}}{\partial t} & = & 0 \label{eq:max4} \\ 
            \square a + \frac{m^{2}_{a} c^4}{\hbar^2} a & = & \frac{g_{a\gamma\gamma}}{\mu_0 c}\mathbf{E}\cdot\mathbf{B} \label{eq:max5}
	\end{eqnarray}

We solve these equations in the background of an axion field
\begin{equation}
a(t) \, = \, a_0 \, \cos\left(\omega_{a} t + \phi_0\right),
\end{equation}
where $\phi_0$ is an arbitrary phase and $\omega_a = m_a c^2/\hbar$.  Within the solar system, we expect that the field amplitude $a_0$
 is given by
 \begin{equation}
 \frac{1}{2 \hbar^2} m_a^2 \langle |a_0|^2 \rangle \approx \rho_{\text{DM}},
 \end{equation}
 where $\rho_{\text{DM}} \approx 0.4~\text{GeV}/c^2/\text{cm}^3$ is the expected local dark matter mass density and the amplitude is averaged over many coherence times.  If we apply a magnetic field $\mathbf{B}_0$ over a region of typical size $L$ with $L m_a v/\hbar \ll 1$ then, as shown in a companion paper~\cite{Berger:2023forthcoming}, the generated electric field will be
\begin{equation}\label{eq:axionEField}
    \mathbf{E} = - c g_{a\gamma\gamma} a \mathbf{B}_0 = - \hbar c g_{a\gamma\gamma} \frac{\sqrt{2 \rho_{\text{DM}}}}{m_a} \mathbf{B}_0.
\end{equation}
There is an additional constant field depending on the phase of the axion when the magnetic field is turned on, but this does not affect the amplitude of the oscillating $E$-field~(see \cite{Berger:2023forthcoming} for details).  For typical configurations and ALP models, we expect
\begin{equation}
     |\mathbf{E}| \approx 74~\text{kV}/\text{m} \times \frac{g_{a\gamma\gamma}/\sqrt{\hbar c}}{10^{-10}~\text{GeV}^{-1}}\cdot  \frac{10^{-21}~\text{eV}}{m_a c^2} \cdot \frac{|\mathbf{B}_0|}{1~\text{mT}}.
\end{equation}
Validity of these calculations further requires that
\begin{equation}
    \hbar \epsilon_0^{1/2} \frac{g_{a\gamma\gamma} |\mathbf{B}_0|}{m_a} \ll 1,
\end{equation}
or
\begin{equation}
    \frac{g_{a\gamma\gamma}/\sqrt{\hbar c}}{10^{-9}~\text{GeV}^{-1}} \cdot  \frac{2 \times 10^{-20}~\text{eV}}{m_a c^2} \cdot \frac{|\mathbf{B}_0|}{0.1~\text{mT}} \ll 1.
\end{equation}

In addition to this electric field, there is a generated magnetic field, but it is expected to be suppressed by $v/c \sim 10^{-3}$, where $v$ is the local velocity dispersion of dark matter.  The field can be written, to leading order in $v/c$, as
\begin{equation}\label{eq:axionBField}
    \mathbf{B} = \frac{1}{c^2} \mathbf{v} \times \mathbf{E},
\end{equation}
which has a magnitude
\begin{equation}\label{eq:numericalB}
    \frac{|\mathbf{B}|}{|\mathbf{B}_0|} \approx 2.4 \cdot 10^{-4} \frac{g_{a\gamma\gamma}/\sqrt{\hbar c}}{10^{-10}~\text{GeV}^{-1}}\cdot  \frac{10^{-21}~\text{eV}}{m_a c^2}.
\end{equation}
Note that the velocity here will not have a magnitude exactly given by the dispersion, but rather sampled randomly as described below. 

For ALPs in the mass range of interest in this work traditional techniques typically employed in searches for the QCD axion, such as resonant conversion to photons in a cavity \cite{Hagmann1990,Boutan2018,HAYSTAC:2020kwv,Crisosto2020}, becomes prohibitive as the latter would have to be comparable in size to a small galaxy. Instead, one must rely on alternative signatures such as the axion sourced electric and magnetic field. Experiments such as ABRACADABRA \cite{Ouellet:2018beu} attempt to detect the small axion sourced magnetic field using the exquisite sensitivity of magnometers such as Superconducting Quantum Interference Devices (SQUIDs). Comparatively little attention has been paid to the electric field, a gap we attempt to fill in this work. 

{\bf Effect on atoms.} The axion sourced electric field generates a Stark shift in the transition frequencies of atomic excitations. High precision atomic spectroscopy is a well-suited tool for detecting ultralight axions, as the Stark shift leads to a time-varying shift in the measured frequency of atomic transitions in the presence of a static background magnetic field. The variation is periodic, with a frequency set by the rest energy of the dark photons to be of order $10^{-6}~\text{Hz}$ for a mass of $10^{-21}~\text{eV}/c^2$. The amplitude, frequency and velocity of the axion are all stochastic and vary on a coherence time $\tau_{c} \sim \hbar / (m_{a} v^{2}) \sim 10^{6}/\omega_a$ \cite{Foster:2017hbq,Lisanti:2021vij}.
While the frequency only varies by a small amount, the amplitude and velocity vary significantly.  The velocity is only particularly relevant for the response magnetic field, which leads to a much smaller effect, while variations in the amplitude leads to an effective reduction of 2.7 in the sensitivity to the electric field \cite{Centers:2019dyn}.

For the axion mass range of interest, the induced electric and magnetic fields vary slowly. Such a slow variation has previously been searched for using spectroscopic measurements of the $\text{Al}^+$, Sr, and Yb optical clocks in \cite{2020arXiv200514694O}. With no significant variation found, the authors of \cite{2020arXiv200514694O} established model independent constraints on the time varying shift on the various optical transition frequencies in question and apply these to the dilaton field, an ultralight spin-0 dark matter candidate. It couples directly to $\mathbf{E}^2 - \mathbf{B}^2$ (unlike the axion, which couples to $\mathbf{E}\cdot\mathbf{B}$) and predicts a variation of the fine structure constant $\alpha$ with time \cite{VanTilburg:2015oza, PhysRevLett.117.061301,2020arXiv200514694O}. Similarly, we find that since a bias static magnetic field of order 0.1 mT was applied for each of these recent spectroscopic experiments, the results, with systematic uncertainties below $10^{-18}$~\cite{2020arXiv200514694O}, are sufficient to set leading laboratory sensitivity to ultralight ALPs. These are complementary to constraints using the cosmic birefringence effect on the Cosmic Microwave Background (CMB) \cite{Fedderke:2019ajk,PhysRevD.105.022006,SPT-3G:2022ods} and polarimetric studies of astrophysical systems such as observations of the Crab Supernova remnant \cite{Ivanov:2018byi} and galactic pulsars observed by the Parkes Pulsar Timing Array \cite{Castillo:2022zfl}. It is important to stress here that while both the dilaton and ALPs cause a time varying shift in the transition frequencies measured in these spectroscopic measurements, the mechanisms are very different. The dilaton induces a time varying shift in the value of the fine structure constant, an effect that is known and has been searched for using atomic systems, for example in \cite{Hees:2016gop,VanTilburg:2015oza}, or using other methods like quasar absorption spectra such as in \cite{2015MNRAS.454.3082W}. ALPs, on the other hand, induce direct time varying Stark and Zeeman shifts in the transition frequencies of these spectroscopic measurements.  

We first focus on the significantly larger Stark shift. 
 If the fine- or hyperfine-structure splittings were sufficiently small, then the Stark shift would be linear in the axion-induced electric field. However, as this is not the case here, the leading effect is a quadratic Stark shift seen at second order in perturbation theory. The $N$th atomic state gets a correction given by
\begin{equation}
    \Delta E_N = \sum_{K \neq N} \frac{\langle N | e \mathbf{E} \cdot \mathbf{x} |K \rangle \langle K |e \mathbf{E} \cdot \mathbf{x} | N\rangle}{E_N - E_K}.
\end{equation}
This sum is directly related to the static electric polarizability $\alpha_D$ (in atomic units, with explicit conversion factors included) of the atomic state $|N\rangle$, by
\begin{equation}
    \alpha_D = -\frac{1}{\epsilon_0 a_B^3} \, \sum_{K \neq N} \frac{\langle N | e z |K \rangle \langle K | e z | N\rangle}{E_N - E_K},
\end{equation}
where explicit factors of the electric charge $e$, Bohr radius $a_B$ and Hartree energy $E_h = \hbar^2 /(m_e a_B^2)$ are included to convert the result into SI units. Using this polarizability we can write
\begin{equation}\label{eq:EStark}
    \Delta E_N = - \alpha_D \, a_B \, \kappa \,\mathbf{B}^{2}_{0}  \cos^{2}\left(\omega_a t + \phi_{0}\right) \, ,
\end{equation}
where we define
\begin{equation}
   \kappa = \frac{4 \pi   \hbar^2 \, c^2  \, \epsilon_0 \, a_B^2 \, \rho_{\text{DM}} \, g^{2}_{a\gamma\gamma}}{m_a^2} 
\end{equation}

{\bf Limits from current measurements.} To apply this calculation to the recent  measurements~\cite{2020arXiv200514694O} of the ratio of transition frequencies $\nu_1$ and $\nu_2$, $R = \nu_1 / \nu_2$, between pairs of atomic or ionic systems 1 and 2 referring to $\text{Al}^+/\text{Sr}$, $\text{Al}^+/\text{Yb}$, and $\text{Yb}/\text{Sr}$. The relevant quantities for each transition are listed in Table \ref{tab:atomic-transition}. The ground state of all these atoms is the $^{1}$S$_{0}$ state, which features two valence electrons in an s shell with total angular momentum zero while the excited state is $^{1}$P$_{0}$, in which one of the s shell electrons is excited to a p shell and the total angular momentum is one. These states are especially chosen because of their insensitivity to external perturbations, such as electric and magnetic fields, since a major goal of these experiments is to achieve a fractional uncertainty below 10$^{-18}$, one of the criteria for the redefinition of the second \cite{Riehle_2018}. As such, Stark and Zeeman shifts are considered unwanted background that need to be reduced as much as possible. To eliminate first order Zeeman shifts, which are proportional to $m_{F}$ (where F is the sum of electron and nuclear spins and $m_{F}$ its projection along the z axis), the eigenstates considered are those with $m_{F}$ = 0. This is done by averaging the highest and lowest $m_{F}$ states for each atom, such that the Zeeman effects precisely cancel at first order. This cancellation helps ensure that the leading shift will indeed be due to the induced electric field.

The time series data of these frequency ratio measurements was fit to a sinusoidal curve and a 95$\%$ confidence interval bound was obtained on the amplitude $\delta_{R}$ of that sinusoid as a function of its frequency of variation in Figures 4 a-c of \cite{2020arXiv200514694O}. We can determine this amplitude as
\begin{equation}
    \delta_R = \text{max} \frac{R(t) - \overline{R}}{\overline{R}},
\end{equation}
where $\overline{R}$ is the average ratio $R$. Since these 95$\%$ confidence interval bounds are model independent, we can recast them in a form appropriate for the dark photon model, for which we find the amplitude of the oscillation of $R$ due to the Stark shift to be
\begin{equation}
    \delta_R = \kappa \,  \left|\frac{\Delta \alpha_{D,2}\mathbf{B}^{2}_{02}}{\hbar \, \omega_2} - \frac{\Delta\alpha_{D,1}\mathbf{B}^{2}_{01}}{\hbar \, \omega_1}\right|,
\end{equation}
where $i= 1,2$ refers to the atomic system being probed, $B_{0i}$, is the applied static magnetic field for that system, and $\Delta \alpha_{D,i}$ are the differential static dipole polarizabilities between the two states of the transition. We assume the ratio is probed at approximately same probe time. 
Note that the frequency of this oscillation is given by
\begin{equation}
    \frac{2 \, m_{a} \, c^2}{h} = 4.8 \cdot 10^{-6}~\text{Hz} \, \frac{m_{a} \, c^2}{10^{-20}~\text{eV}},
\end{equation}
which is a factor of 2 larger than that of the field itself since the observed effect, being second order in perturbation theory, goes like the square of the field. 

\begin{table*}[!t]
\begin{tabular}{c c c c c c } 
\hline\hline
Atom & Transition & Energy (eV) & Ground state $\alpha_D$ (a.u.) & Excited state $\alpha_D$ (a.u.) & $B_{0}$ (mT)\\
\hline
$\text{Al}^+$ & $3s^2~{}^1S_0$ --- $3s3p~{}^3P_0$ & $4.643$ & 23.780 & 24.175 & 0.12 \\
$\text{Sr}$ & $5s^2~{}^1S_0$ --- $5s5p~{}^3P_0$ & $1.776$ & 193 & 410 & 5.7 $\times$ 10$^{-2}$ \\
$\text{Yb}$ & $4f^{14}6s^2~{}^1S_{0}$ --- $4f^{14}6s6p~{}^3P_{0}$ & $2.145$  & 139 & 257 & 0.1 \\
\hline\hline
\end{tabular}
\caption{Atomic transitions considered by~\cite{2020arXiv200514694O}. The transition energies are taken from \cite{2020arXiv200514694O}. Polarizability data is taken from~\cite{PhysRevLett.107.143006,PhysRevLett.123.033201} ($\text{Al}^{+}$) and \cite{2010JPhB...43m5004G} (Yb and Sr). Values for the applied bias magnetic field are taken from \cite{PhysRevLett.123.033201} for Al$^{+}$, \cite{2019Metro..56f5004B} for Sr, and \cite{McGrew:2018mqk} for Yb. }\label{tab:atomic-transition}
\end{table*}

{\bf Axion sourced magnetic field.} From equation \eqref{eq:axionBField}, we note that there is also a Zeeman shift due to the axion sourced magnetic field, which can be distinguished from the static applied $B_{0}\hat{z}$ field due to its time varying nature. Despite being suppressed by a factor of the dark matter velocity $\langle v \rangle \approx 10^{-3} c$, this Zeeman shift is is potentially first order and can hence dominate over the Stark shift, which is second order. However, as already discussed, the Zeeman shifts for the states of interest in this work are all second order in perturbation theory, which makes them suppressed compared to the Stark shift by a factor of $10^{-6}$. To get a quantitative estimate, we estimate the former as    
\begin{equation}
    \Delta E_{\text{Z}} \simeq \frac{\mu_{B}^{2}\mathbf{B}^{2}}{\Delta E_{0}},
\end{equation}
where $\Delta E_{0}$ is the smallest energy difference between the unperturbed eigenstates. For the optical clocks used in this work, $\Delta E_{0} \sim 1~\text{eV}$, we can apply the result \eqref{eq:numericalB} to find
\begin{equation}
    \frac{\Delta E_{\text{Z}}}{\text{eV}} \simeq 1.9 \cdot 10^{-22} \ \left(\frac{B_{0}}{\text{mT}} \ \frac{g_{a\gamma\gamma}/\sqrt{\hbar c}}{10^{-10} \ \text{GeV}^{-1}}\frac{10^{-21} \ \text{eV}}{m_{a} c^2}\right)^{2} .
\end{equation}
Comparing this to an estimate of the Stark induced correction in~\eqref{eq:EStark},
\begin{equation}
    \frac{\Delta E_{\text{S}}}{\text{eV}} \simeq  1.4 \cdot 10^{-13}\, \alpha_{D}\, \left(\frac{B_{0}}{ \text{mT}}~\frac{g_{a\gamma\gamma}/\sqrt{\hbar c}}{10^{-10} \ \text{GeV}^{-1}}\frac{10^{-21}~\text{eV}}{m_{a}} \right)^{2},
\end{equation}
we see that the Stark shift is the dominant effect. 

\begin{figure}[tbh!]
    \centering
        \includegraphics[width=0.47\textwidth]{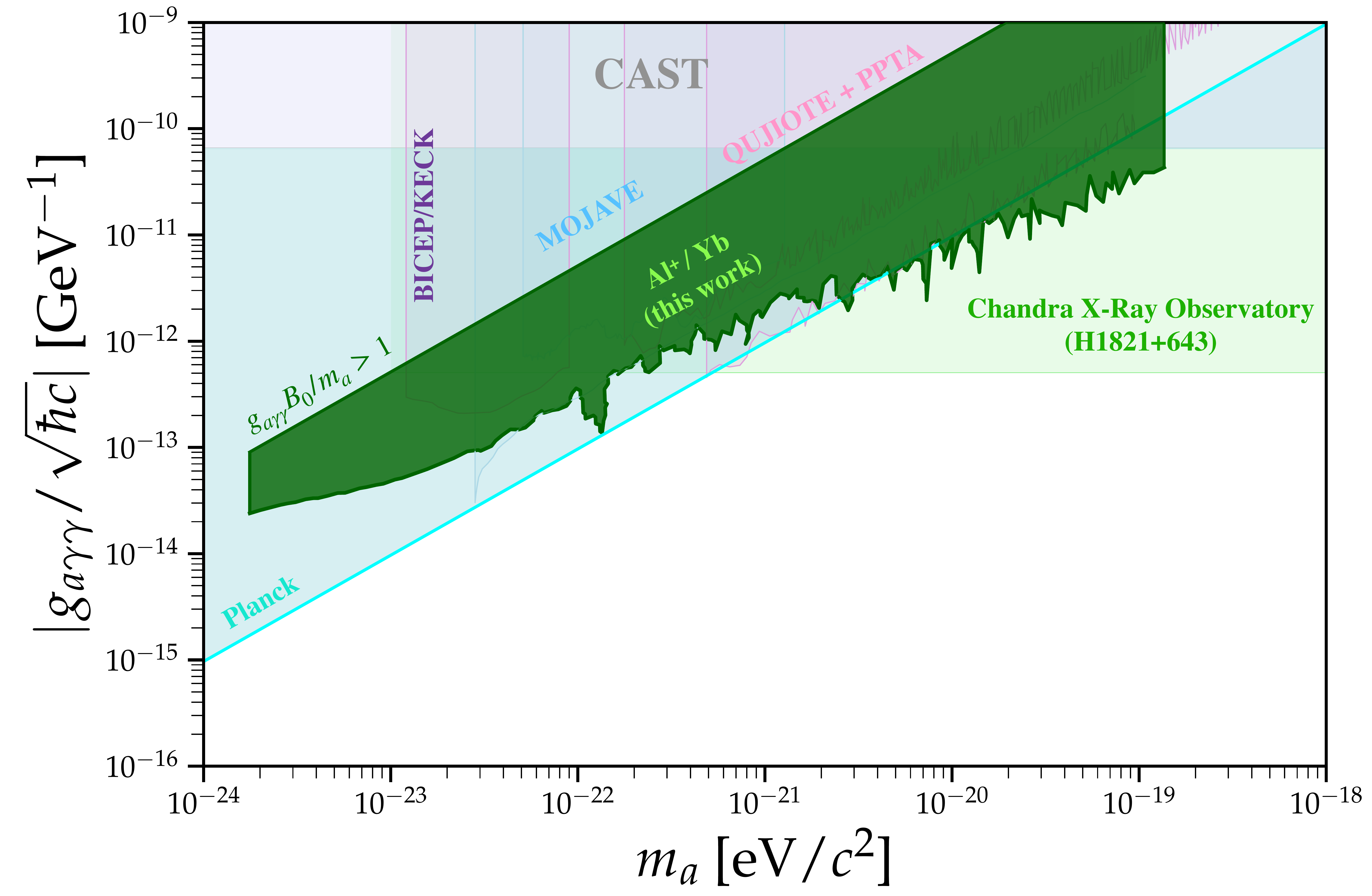}
    \caption{ Bounds on the ultralight ALPs comprising all the dark matter from precision atomic spectroscopy (dark green). The dominant bound comes from deviations in the ${\text{Al}^{+}}/\text{Yb}$ frequency ratio.  
    At large $g_{a\gamma\gamma}$, we truncate our bound as our perturbative calculation breaks down at $B_0 = 0.1~\text{mT}$.
    We additionally show astrophysical bounds from CAST~\cite{CAST:2017uph}, Planck~\cite{Fedderke:2019ajk}, BICEP/KECK~\cite{PhysRevD.105.022006}, MOJAVE~\cite{Ivanov:2018byi}, Parkes Pulsar Timing Array~\cite{Castillo:2022zfl}, and leading constraints from Chandra~\cite{2022MNRAS.510.1264S}.  The astrophysical bounds are constructed with the \texttt{AxionLimits} package~\cite{ciaran_o_hare_2020_3932430}.} \label{fig:boundsfit}
    
\end{figure}

{\bf Results.} The constraints we obtain using the dominant Stark shift are shown in Fig.~\ref{fig:boundsfit}. The mass range is from 10$^{-23}$ eV $ < m_{a} < $ 10$^{-18}$ eV, which is the corresponding frequency range over which the fit was performed in \cite{2020arXiv200514694O}. The upper end of 10$^{-18}$ eV, or frequency of $10^{-2}$ Hz, was chosen by the authors as it corresponds to a conservative minimum observable period well above the experiment's servo time constant of roughly 10s, which sets how frequently the transitions in question can be probed. The lower end of 10$^{-23}$ eV corresponds to roughly a year, the period over which data was taken.  

{\bf Future directions.} Progress continues to be made on probing atomic systems ever more precisely, opening the question of what future measurements could improve on the constraints demonstrated in this work. Unfortunately, while Stark and Zeeman shifts represent a signal for ultralight ALP searches, they are also precisely systematic uncertainties that these experiments attempt to mitigate. 

For ALPs, Stark and Zeeman shifts may also be a distinguishing signature for discovery as the latter, being sourced by the same field, are correlated. Such a correlation could be distinguished from other dark matter candidates such as the dilaton or dark photons as they depend on the externally applied magnetic field in two ways: firstly, the axion sourced external electric field is parallel to the externally applied magnetic field, and secondly, both the Stark and Zeeman shifts are proportional to the externally applied magnetic field. Hence, varying the strength and direction of the external magnetic field may provide a crucial signature for the axion sourced, correlated, Stark and Zeeman shifts. 

Another possibility is the detection of ultralight dark matter induced excess motion (EMM) in with single ion clocks such as that of Al$^{+}$ in \cite{2020arXiv200514694O}. Techniques such as sideband spectroscopy can measure EMM with a precision of $\Delta\nu / \nu \simeq$ 10$^{-18}$ \cite{PhysRevLett.123.033201}. Since this effect is first order in the electric field, EMM could provide unprecedented sensitivity to ultralight ALPs. \\

{\bf Acknowledgments.} We thank Masha Baryakhtar, Sam Brewer, Christian Sanner, Joseph Bramante, Michael Fedderke, Saarik Kalia, and Ningqiang Song for useful discussions. This material is based upon work supported by the National Science Foundation under Grant No.\ 2112789.

\bibliographystyle{apsrev4-2}
\bibliography{dpdm}

\begin{thebibliography}{43}%
\makeatletter
\providecommand \@ifxundefined [1]{%
 \@ifx{#1\undefined}
}%
\providecommand \@ifnum [1]{%
 \ifnum #1\expandafter \@firstoftwo
 \else \expandafter \@secondoftwo
 \fi
}%
\providecommand \@ifx [1]{%
 \ifx #1\expandafter \@firstoftwo
 \else \expandafter \@secondoftwo
 \fi
}%
\providecommand \natexlab [1]{#1}%
\providecommand \enquote  [1]{``#1''}%
\providecommand \bibnamefont  [1]{#1}%
\providecommand \bibfnamefont [1]{#1}%
\providecommand \citenamefont [1]{#1}%
\providecommand \href@noop [0]{\@secondoftwo}%
\providecommand \href [0]{\begingroup \@sanitize@url \@href}%
\providecommand \@href[1]{\@@startlink{#1}\@@href}%
\providecommand \@@href[1]{\endgroup#1\@@endlink}%
\providecommand \@sanitize@url [0]{\catcode `\\12\catcode `\$12\catcode
  `\&12\catcode `\#12\catcode `\^12\catcode `\_12\catcode `\%12\relax}%
\providecommand \@@startlink[1]{}%
\providecommand \@@endlink[0]{}%
\providecommand \url  [0]{\begingroup\@sanitize@url \@url }%
\providecommand \@url [1]{\endgroup\@href {#1}{\urlprefix }}%
\providecommand \urlprefix  [0]{URL }%
\providecommand \Eprint [0]{\href }%
\providecommand \doibase [0]{https://doi.org/}%
\providecommand \selectlanguage [0]{\@gobble}%
\providecommand \bibinfo  [0]{\@secondoftwo}%
\providecommand \bibfield  [0]{\@secondoftwo}%
\providecommand \translation [1]{[#1]}%
\providecommand \BibitemOpen [0]{}%
\providecommand \bibitemStop [0]{}%
\providecommand \bibitemNoStop [0]{.\EOS\space}%
\providecommand \EOS [0]{\spacefactor3000\relax}%
\providecommand \BibitemShut  [1]{\csname bibitem#1\endcsname}%
\let\auto@bib@innerbib\@empty
\bibitem [{\citenamefont {Peccei}\ and\ \citenamefont
  {Quinn}(1977)}]{Peccei:1977hh}%
  \BibitemOpen
  \bibfield  {author} {\bibinfo {author} {\bibfnamefont {R.~D.}\ \bibnamefont
  {Peccei}}\ and\ \bibinfo {author} {\bibfnamefont {H.~R.}\ \bibnamefont
  {Quinn}},\ }\href {https://doi.org/10.1103/PhysRevLett.38.1440} {\bibfield
  {journal} {\bibinfo  {journal} {Phys. Rev. Lett.}\ }\textbf {\bibinfo
  {volume} {38}},\ \bibinfo {pages} {1440} (\bibinfo {year}
  {1977})}\BibitemShut {NoStop}%
\bibitem [{\citenamefont {Wilczek}(1978)}]{Wilczek:1977pj}%
  \BibitemOpen
  \bibfield  {author} {\bibinfo {author} {\bibfnamefont {F.}~\bibnamefont
  {Wilczek}},\ }\href {https://doi.org/10.1103/PhysRevLett.40.279} {\bibfield
  {journal} {\bibinfo  {journal} {Phys. Rev. Lett.}\ }\textbf {\bibinfo
  {volume} {40}},\ \bibinfo {pages} {279} (\bibinfo {year} {1978})}\BibitemShut
  {NoStop}%
\bibitem [{\citenamefont {Weinberg}(1978)}]{Weinberg:1977ma}%
  \BibitemOpen
  \bibfield  {author} {\bibinfo {author} {\bibfnamefont {S.}~\bibnamefont
  {Weinberg}},\ }\href {https://doi.org/10.1103/PhysRevLett.40.223} {\bibfield
  {journal} {\bibinfo  {journal} {Phys. Rev. Lett.}\ }\textbf {\bibinfo
  {volume} {40}},\ \bibinfo {pages} {223} (\bibinfo {year} {1978})}\BibitemShut
  {NoStop}%
\bibitem [{\citenamefont {Kim}(1979)}]{Kim:1979if}%
  \BibitemOpen
  \bibfield  {author} {\bibinfo {author} {\bibfnamefont {J.~E.}\ \bibnamefont
  {Kim}},\ }\href {https://doi.org/10.1103/PhysRevLett.43.103} {\bibfield
  {journal} {\bibinfo  {journal} {Phys. Rev. Lett.}\ }\textbf {\bibinfo
  {volume} {43}},\ \bibinfo {pages} {103} (\bibinfo {year} {1979})}\BibitemShut
  {NoStop}%
\bibitem [{\citenamefont {Shifman}\ \emph {et~al.}(1980)\citenamefont
  {Shifman}, \citenamefont {Vainshtein},\ and\ \citenamefont
  {Zakharov}}]{Shifman:1979if}%
  \BibitemOpen
  \bibfield  {author} {\bibinfo {author} {\bibfnamefont {M.~A.}\ \bibnamefont
  {Shifman}}, \bibinfo {author} {\bibfnamefont {A.~I.}\ \bibnamefont
  {Vainshtein}},\ and\ \bibinfo {author} {\bibfnamefont {V.~I.}\ \bibnamefont
  {Zakharov}},\ }\href {https://doi.org/10.1016/0550-3213(80)90209-6}
  {\bibfield  {journal} {\bibinfo  {journal} {Nucl. Phys. B}\ }\textbf
  {\bibinfo {volume} {166}},\ \bibinfo {pages} {493} (\bibinfo {year}
  {1980})}\BibitemShut {NoStop}%
\bibitem [{\citenamefont {Dine}\ \emph {et~al.}(1981)\citenamefont {Dine},
  \citenamefont {Fischler},\ and\ \citenamefont {Srednicki}}]{Dine:1981rt}%
  \BibitemOpen
  \bibfield  {author} {\bibinfo {author} {\bibfnamefont {M.}~\bibnamefont
  {Dine}}, \bibinfo {author} {\bibfnamefont {W.}~\bibnamefont {Fischler}},\
  and\ \bibinfo {author} {\bibfnamefont {M.}~\bibnamefont {Srednicki}},\ }\href
  {https://doi.org/10.1016/0370-2693(81)90590-6} {\bibfield  {journal}
  {\bibinfo  {journal} {Phys. Lett. B}\ }\textbf {\bibinfo {volume} {104}},\
  \bibinfo {pages} {199} (\bibinfo {year} {1981})}\BibitemShut {NoStop}%
\bibitem [{\citenamefont {Zhitnitsky}(1980)}]{Zhitnitsky:1980tq}%
  \BibitemOpen
  \bibfield  {author} {\bibinfo {author} {\bibfnamefont {A.~R.}\ \bibnamefont
  {Zhitnitsky}},\ }\href@noop {} {\bibfield  {journal} {\bibinfo  {journal}
  {Sov. J. Nucl. Phys.}\ }\textbf {\bibinfo {volume} {31}},\ \bibinfo {pages}
  {260} (\bibinfo {year} {1980})}\BibitemShut {NoStop}%
\bibitem [{\citenamefont {Preskill}\ \emph {et~al.}(1983)\citenamefont
  {Preskill}, \citenamefont {Wise},\ and\ \citenamefont
  {Wilczek}}]{Preskill:1982cy}%
  \BibitemOpen
  \bibfield  {author} {\bibinfo {author} {\bibfnamefont {J.}~\bibnamefont
  {Preskill}}, \bibinfo {author} {\bibfnamefont {M.~B.}\ \bibnamefont {Wise}},\
  and\ \bibinfo {author} {\bibfnamefont {F.}~\bibnamefont {Wilczek}},\ }\href
  {https://doi.org/10.1016/0370-2693(83)90637-8} {\bibfield  {journal}
  {\bibinfo  {journal} {Phys. Lett. B}\ }\textbf {\bibinfo {volume} {120}},\
  \bibinfo {pages} {127} (\bibinfo {year} {1983})}\BibitemShut {NoStop}%
\bibitem [{\citenamefont {Abbott}\ and\ \citenamefont
  {Sikivie}(1983)}]{Abbott:1982af}%
  \BibitemOpen
  \bibfield  {author} {\bibinfo {author} {\bibfnamefont {L.~F.}\ \bibnamefont
  {Abbott}}\ and\ \bibinfo {author} {\bibfnamefont {P.}~\bibnamefont
  {Sikivie}},\ }\href {https://doi.org/10.1016/0370-2693(83)90638-X} {\bibfield
   {journal} {\bibinfo  {journal} {Phys. Lett. B}\ }\textbf {\bibinfo {volume}
  {120}},\ \bibinfo {pages} {133} (\bibinfo {year} {1983})}\BibitemShut
  {NoStop}%
\bibitem [{\citenamefont {Dine}\ and\ \citenamefont
  {Fischler}(1983)}]{Dine:1982ah}%
  \BibitemOpen
  \bibfield  {author} {\bibinfo {author} {\bibfnamefont {M.}~\bibnamefont
  {Dine}}\ and\ \bibinfo {author} {\bibfnamefont {W.}~\bibnamefont
  {Fischler}},\ }\href {https://doi.org/10.1016/0370-2693(83)90639-1}
  {\bibfield  {journal} {\bibinfo  {journal} {Phys. Lett. B}\ }\textbf
  {\bibinfo {volume} {120}},\ \bibinfo {pages} {137} (\bibinfo {year}
  {1983})}\BibitemShut {NoStop}%
\bibitem [{\citenamefont {{Moore}}(1994)}]{1994Natur.370..629M}%
  \BibitemOpen
  \bibfield  {author} {\bibinfo {author} {\bibfnamefont {B.}~\bibnamefont
  {{Moore}}},\ }\href {https://doi.org/10.1038/370629a0} {\bibfield  {journal}
  {\bibinfo  {journal} {\nat}\ }\textbf {\bibinfo {volume} {370}},\ \bibinfo
  {pages} {629} (\bibinfo {year} {1994})}\BibitemShut {NoStop}%
\bibitem [{\citenamefont {{Boylan-Kolchin}}\ \emph {et~al.}(2011)\citenamefont
  {{Boylan-Kolchin}}, \citenamefont {{Bullock}},\ and\ \citenamefont
  {{Kaplinghat}}}]{2011MNRAS.415L..40B}%
  \BibitemOpen
  \bibfield  {author} {\bibinfo {author} {\bibfnamefont {M.}~\bibnamefont
  {{Boylan-Kolchin}}}, \bibinfo {author} {\bibfnamefont {J.~S.}\ \bibnamefont
  {{Bullock}}},\ and\ \bibinfo {author} {\bibfnamefont {M.}~\bibnamefont
  {{Kaplinghat}}},\ }\href {https://doi.org/10.1111/j.1745-3933.2011.01074.x}
  {\bibfield  {journal} {\bibinfo  {journal} {Mon. Notices Royal Astron. Soc}\
  }\textbf {\bibinfo {volume} {415}},\ \bibinfo {pages} {L40} (\bibinfo {year}
  {2011})},\ \Eprint {https://arxiv.org/abs/1103.0007} {arXiv:1103.0007
  [astro-ph.CO]} \BibitemShut {NoStop}%
\bibitem [{\citenamefont {{Kauffmann}}\ \emph {et~al.}(1993)\citenamefont
  {{Kauffmann}}, \citenamefont {{White}},\ and\ \citenamefont
  {{Guiderdoni}}}]{1993MNRAS.264..201K}%
  \BibitemOpen
  \bibfield  {author} {\bibinfo {author} {\bibfnamefont {G.}~\bibnamefont
  {{Kauffmann}}}, \bibinfo {author} {\bibfnamefont {S.~D.~M.}\ \bibnamefont
  {{White}}},\ and\ \bibinfo {author} {\bibfnamefont {B.}~\bibnamefont
  {{Guiderdoni}}},\ }\href {https://doi.org/10.1093/mnras/264.1.201} {\bibfield
   {journal} {\bibinfo  {journal} {MNRAS}\ }\textbf {\bibinfo {volume} {264}},\
  \bibinfo {pages} {201} (\bibinfo {year} {1993})}\BibitemShut {NoStop}%
\bibitem [{\citenamefont {{Klypin}}\ \emph {et~al.}(1999)\citenamefont
  {{Klypin}}, \citenamefont {{Kravtsov}}, \citenamefont {{Valenzuela}},\ and\
  \citenamefont {{Prada}}}]{1999ApJ...522...82K}%
  \BibitemOpen
  \bibfield  {author} {\bibinfo {author} {\bibfnamefont {A.}~\bibnamefont
  {{Klypin}}}, \bibinfo {author} {\bibfnamefont {A.~V.}\ \bibnamefont
  {{Kravtsov}}}, \bibinfo {author} {\bibfnamefont {O.}~\bibnamefont
  {{Valenzuela}}},\ and\ \bibinfo {author} {\bibfnamefont {F.}~\bibnamefont
  {{Prada}}},\ }\href {https://doi.org/10.1086/307643} {\bibfield  {journal}
  {\bibinfo  {journal} {\apj}\ }\textbf {\bibinfo {volume} {522}},\ \bibinfo
  {pages} {82} (\bibinfo {year} {1999})},\ \Eprint
  {https://arxiv.org/abs/astro-ph/9901240} {arXiv:astro-ph/9901240 [astro-ph]}
  \BibitemShut {NoStop}%
\bibitem [{\citenamefont {Wilczek}(1987)}]{PhysRevLett.58.1799}%
  \BibitemOpen
  \bibfield  {author} {\bibinfo {author} {\bibfnamefont {F.}~\bibnamefont
  {Wilczek}},\ }\href {https://doi.org/10.1103/PhysRevLett.58.1799} {\bibfield
  {journal} {\bibinfo  {journal} {Phys. Rev. Lett.}\ }\textbf {\bibinfo
  {volume} {58}},\ \bibinfo {pages} {1799} (\bibinfo {year}
  {1987})}\BibitemShut {NoStop}%
\bibitem [{\citenamefont {Berger}\ and\ \citenamefont
  {Bhoonah}()}]{Berger:2023forthcoming}%
  \BibitemOpen
  \bibfield  {author} {\bibinfo {author} {\bibfnamefont {J.}~\bibnamefont
  {Berger}}\ and\ \bibinfo {author} {\bibfnamefont {A.}~\bibnamefont
  {Bhoonah}},\ }\href@noop {} {\ }\Eprint {https://arxiv.org/abs/to appear} {to
  appear} \BibitemShut {NoStop}%
\bibitem [{\citenamefont {Hagmann}\ \emph {et~al.}(1990)\citenamefont
  {Hagmann}, \citenamefont {Sikivie}, \citenamefont {Sullivan},\ and\
  \citenamefont {Tanner}}]{Hagmann1990}%
  \BibitemOpen
  \bibfield  {author} {\bibinfo {author} {\bibfnamefont {C.}~\bibnamefont
  {Hagmann}}, \bibinfo {author} {\bibfnamefont {P.}~\bibnamefont {Sikivie}},
  \bibinfo {author} {\bibfnamefont {N.~S.}\ \bibnamefont {Sullivan}},\ and\
  \bibinfo {author} {\bibfnamefont {D.~B.}\ \bibnamefont {Tanner}},\ }\href
  {https://doi.org/10.1103/PhysRevD.42.1297} {\bibfield  {journal} {\bibinfo
  {journal} {Phys. Rev. D}\ }\textbf {\bibinfo {volume} {42}},\ \bibinfo
  {pages} {1297} (\bibinfo {year} {1990})}\BibitemShut {NoStop}%
\bibitem [{\citenamefont {Boutan}\ \emph {et~al.}(2018)\citenamefont {Boutan},
  \citenamefont {Jones}, \citenamefont {LaRoque}, \citenamefont {Oblath},
  \citenamefont {Cervantes}, \citenamefont {Du}, \citenamefont {Force},
  \citenamefont {Kimes}, \citenamefont {Ottens}, \citenamefont {Rosenberg},
  \citenamefont {Rybka}, \citenamefont {Yang}, \citenamefont {Carosi},
  \citenamefont {Woollett}, \citenamefont {Bowring}, \citenamefont {Chou},
  \citenamefont {Khatiwada}, \citenamefont {Sonnenschein}, \citenamefont
  {Wester}, \citenamefont {Bradley}, \citenamefont {Daw}, \citenamefont
  {Agrawal}, \citenamefont {Dixit}, \citenamefont {Clarke}, \citenamefont
  {O'Kelley}, \citenamefont {Crisosto}, \citenamefont {Gleason}, \citenamefont
  {Jois}, \citenamefont {Sikivie}, \citenamefont {Stern}, \citenamefont
  {Sullivan}, \citenamefont {Tanner}, \citenamefont {Harrington},\ and\
  \citenamefont {Lentz}}]{Boutan2018}%
  \BibitemOpen
  \bibfield  {author} {\bibinfo {author} {\bibfnamefont {C.}~\bibnamefont
  {Boutan}}, \bibinfo {author} {\bibfnamefont {M.}~\bibnamefont {Jones}},
  \bibinfo {author} {\bibfnamefont {B.~H.}\ \bibnamefont {LaRoque}}, \bibinfo
  {author} {\bibfnamefont {N.~S.}\ \bibnamefont {Oblath}}, \bibinfo {author}
  {\bibfnamefont {R.}~\bibnamefont {Cervantes}}, \bibinfo {author}
  {\bibfnamefont {N.}~\bibnamefont {Du}}, \bibinfo {author} {\bibfnamefont
  {N.}~\bibnamefont {Force}}, \bibinfo {author} {\bibfnamefont
  {S.}~\bibnamefont {Kimes}}, \bibinfo {author} {\bibfnamefont
  {R.}~\bibnamefont {Ottens}}, \bibinfo {author} {\bibfnamefont {L.~J.}\
  \bibnamefont {Rosenberg}}, \bibinfo {author} {\bibfnamefont {G.}~\bibnamefont
  {Rybka}}, \bibinfo {author} {\bibfnamefont {J.}~\bibnamefont {Yang}},
  \bibinfo {author} {\bibfnamefont {G.}~\bibnamefont {Carosi}}, \bibinfo
  {author} {\bibfnamefont {N.}~\bibnamefont {Woollett}}, \bibinfo {author}
  {\bibfnamefont {D.}~\bibnamefont {Bowring}}, \bibinfo {author} {\bibfnamefont
  {A.~S.}\ \bibnamefont {Chou}}, \bibinfo {author} {\bibfnamefont
  {R.}~\bibnamefont {Khatiwada}}, \bibinfo {author} {\bibfnamefont
  {A.}~\bibnamefont {Sonnenschein}}, \bibinfo {author} {\bibfnamefont
  {W.}~\bibnamefont {Wester}}, \bibinfo {author} {\bibfnamefont
  {R.}~\bibnamefont {Bradley}}, \bibinfo {author} {\bibfnamefont {E.~J.}\
  \bibnamefont {Daw}}, \bibinfo {author} {\bibfnamefont {A.}~\bibnamefont
  {Agrawal}}, \bibinfo {author} {\bibfnamefont {A.~V.}\ \bibnamefont {Dixit}},
  \bibinfo {author} {\bibfnamefont {J.}~\bibnamefont {Clarke}}, \bibinfo
  {author} {\bibfnamefont {S.~R.}\ \bibnamefont {O'Kelley}}, \bibinfo {author}
  {\bibfnamefont {N.}~\bibnamefont {Crisosto}}, \bibinfo {author}
  {\bibfnamefont {J.~R.}\ \bibnamefont {Gleason}}, \bibinfo {author}
  {\bibfnamefont {S.}~\bibnamefont {Jois}}, \bibinfo {author} {\bibfnamefont
  {P.}~\bibnamefont {Sikivie}}, \bibinfo {author} {\bibfnamefont
  {I.}~\bibnamefont {Stern}}, \bibinfo {author} {\bibfnamefont {N.~S.}\
  \bibnamefont {Sullivan}}, \bibinfo {author} {\bibfnamefont {D.~B.}\
  \bibnamefont {Tanner}}, \bibinfo {author} {\bibfnamefont {P.~M.}\
  \bibnamefont {Harrington}},\ and\ \bibinfo {author} {\bibfnamefont
  {E.}~\bibnamefont {Lentz}} (\bibinfo {collaboration} {ADMX Collaboration}),\
  }\href {https://doi.org/10.1103/PhysRevLett.121.261302} {\bibfield  {journal}
  {\bibinfo  {journal} {Phys. Rev. Lett.}\ }\textbf {\bibinfo {volume} {121}},\
  \bibinfo {pages} {261302} (\bibinfo {year} {2018})}\BibitemShut {NoStop}%
\bibitem [{\citenamefont {Backes}\ \emph {et~al.}(2021)\citenamefont {Backes}
  \emph {et~al.}}]{HAYSTAC:2020kwv}%
  \BibitemOpen
  \bibfield  {author} {\bibinfo {author} {\bibfnamefont {K.~M.}\ \bibnamefont
  {Backes}} \emph {et~al.} (\bibinfo {collaboration} {HAYSTAC}),\ }\href
  {https://doi.org/10.1038/s41586-021-03226-7} {\bibfield  {journal} {\bibinfo
  {journal} {Nature}\ }\textbf {\bibinfo {volume} {590}},\ \bibinfo {pages}
  {238} (\bibinfo {year} {2021})}\BibitemShut {NoStop}%
\bibitem [{\citenamefont {Crisosto}\ \emph {et~al.}(2020)\citenamefont
  {Crisosto}, \citenamefont {Sikivie}, \citenamefont {Sullivan}, \citenamefont
  {Tanner}, \citenamefont {Yang},\ and\ \citenamefont {Rybka}}]{Crisosto2020}%
  \BibitemOpen
  \bibfield  {author} {\bibinfo {author} {\bibfnamefont {N.}~\bibnamefont
  {Crisosto}}, \bibinfo {author} {\bibfnamefont {P.}~\bibnamefont {Sikivie}},
  \bibinfo {author} {\bibfnamefont {N.~S.}\ \bibnamefont {Sullivan}}, \bibinfo
  {author} {\bibfnamefont {D.~B.}\ \bibnamefont {Tanner}}, \bibinfo {author}
  {\bibfnamefont {J.}~\bibnamefont {Yang}},\ and\ \bibinfo {author}
  {\bibfnamefont {G.}~\bibnamefont {Rybka}},\ }\href
  {https://doi.org/10.1103/PhysRevLett.124.241101} {\bibfield  {journal}
  {\bibinfo  {journal} {Phys. Rev. Lett.}\ }\textbf {\bibinfo {volume} {124}},\
  \bibinfo {pages} {241101} (\bibinfo {year} {2020})}\BibitemShut {NoStop}%
\bibitem [{\citenamefont {Ouellet}\ \emph {et~al.}(2019)\citenamefont {Ouellet}
  \emph {et~al.}}]{Ouellet:2018beu}%
  \BibitemOpen
  \bibfield  {author} {\bibinfo {author} {\bibfnamefont {J.~L.}\ \bibnamefont
  {Ouellet}} \emph {et~al.},\ }\href
  {https://doi.org/10.1103/PhysRevLett.122.121802} {\bibfield  {journal}
  {\bibinfo  {journal} {Phys. Rev. Lett.}\ }\textbf {\bibinfo {volume} {122}},\
  \bibinfo {pages} {121802} (\bibinfo {year} {2019})},\ \Eprint
  {https://arxiv.org/abs/1810.12257} {arXiv:1810.12257 [hep-ex]} \BibitemShut
  {NoStop}%
\bibitem [{\citenamefont {Foster}\ \emph {et~al.}(2018)\citenamefont {Foster},
  \citenamefont {Rodd},\ and\ \citenamefont {Safdi}}]{Foster:2017hbq}%
  \BibitemOpen
  \bibfield  {author} {\bibinfo {author} {\bibfnamefont {J.~W.}\ \bibnamefont
  {Foster}}, \bibinfo {author} {\bibfnamefont {N.~L.}\ \bibnamefont {Rodd}},\
  and\ \bibinfo {author} {\bibfnamefont {B.~R.}\ \bibnamefont {Safdi}},\ }\href
  {https://doi.org/10.1103/PhysRevD.97.123006} {\bibfield  {journal} {\bibinfo
  {journal} {Phys. Rev. D}\ }\textbf {\bibinfo {volume} {97}},\ \bibinfo
  {pages} {123006} (\bibinfo {year} {2018})},\ \Eprint
  {https://arxiv.org/abs/1711.10489} {arXiv:1711.10489 [astro-ph.CO]}
  \BibitemShut {NoStop}%
\bibitem [{\citenamefont {Lisanti}\ \emph {et~al.}(2021)\citenamefont
  {Lisanti}, \citenamefont {Moschella},\ and\ \citenamefont
  {Terrano}}]{Lisanti:2021vij}%
  \BibitemOpen
  \bibfield  {author} {\bibinfo {author} {\bibfnamefont {M.}~\bibnamefont
  {Lisanti}}, \bibinfo {author} {\bibfnamefont {M.}~\bibnamefont {Moschella}},\
  and\ \bibinfo {author} {\bibfnamefont {W.}~\bibnamefont {Terrano}},\ }\href
  {https://doi.org/10.1103/PhysRevD.104.055037} {\bibfield  {journal} {\bibinfo
   {journal} {Phys. Rev. D}\ }\textbf {\bibinfo {volume} {104}},\ \bibinfo
  {pages} {055037} (\bibinfo {year} {2021})},\ \Eprint
  {https://arxiv.org/abs/2107.10260} {arXiv:2107.10260 [astro-ph.CO]}
  \BibitemShut {NoStop}%
\bibitem [{\citenamefont {Centers}\ \emph {et~al.}(2021)\citenamefont {Centers}
  \emph {et~al.}}]{Centers:2019dyn}%
  \BibitemOpen
  \bibfield  {author} {\bibinfo {author} {\bibfnamefont {G.~P.}\ \bibnamefont
  {Centers}} \emph {et~al.},\ }\href
  {https://doi.org/10.1038/s41467-021-27632-7} {\bibfield  {journal} {\bibinfo
  {journal} {Nature Commun.}\ }\textbf {\bibinfo {volume} {12}},\ \bibinfo
  {pages} {7321} (\bibinfo {year} {2021})},\ \Eprint
  {https://arxiv.org/abs/1905.13650} {arXiv:1905.13650 [astro-ph.CO]}
  \BibitemShut {NoStop}%
\bibitem [{\citenamefont {Beloy}\ \emph {et~al.}(2021)\citenamefont {Beloy}
  \emph {et~al.}}]{2020arXiv200514694O}%
  \BibitemOpen
  \bibfield  {author} {\bibinfo {author} {\bibfnamefont {K.}~\bibnamefont
  {Beloy}} \emph {et~al.} (\bibinfo {collaboration} {Boulder Atomic Clock
  Optical Network (BACON) Collaboration}),\ }\href
  {https://doi.org/10.1038/s41586-021-03253-4} {\bibfield  {journal} {\bibinfo
  {journal} {Nature}\ }\textbf {\bibinfo {volume} {591}},\ \bibinfo {pages}
  {564–569} (\bibinfo {year} {2021})}\BibitemShut {NoStop}%
\bibitem [{\citenamefont {Van~Tilburg}\ \emph {et~al.}(2015)\citenamefont
  {Van~Tilburg}, \citenamefont {Leefer}, \citenamefont {Bougas},\ and\
  \citenamefont {Budker}}]{VanTilburg:2015oza}%
  \BibitemOpen
  \bibfield  {author} {\bibinfo {author} {\bibfnamefont {K.}~\bibnamefont
  {Van~Tilburg}}, \bibinfo {author} {\bibfnamefont {N.}~\bibnamefont {Leefer}},
  \bibinfo {author} {\bibfnamefont {L.}~\bibnamefont {Bougas}},\ and\ \bibinfo
  {author} {\bibfnamefont {D.}~\bibnamefont {Budker}},\ }\href
  {https://doi.org/10.1103/PhysRevLett.115.011802} {\bibfield  {journal}
  {\bibinfo  {journal} {Phys. Rev. Lett.}\ }\textbf {\bibinfo {volume} {115}},\
  \bibinfo {pages} {011802} (\bibinfo {year} {2015})},\ \Eprint
  {https://arxiv.org/abs/1503.06886} {arXiv:1503.06886 [physics.atom-ph]}
  \BibitemShut {NoStop}%
\bibitem [{\citenamefont {Hees}\ \emph
  {et~al.}(2016{\natexlab{a}})\citenamefont {Hees}, \citenamefont {Gu\'ena},
  \citenamefont {Abgrall}, \citenamefont {Bize},\ and\ \citenamefont
  {Wolf}}]{PhysRevLett.117.061301}%
  \BibitemOpen
  \bibfield  {author} {\bibinfo {author} {\bibfnamefont {A.}~\bibnamefont
  {Hees}}, \bibinfo {author} {\bibfnamefont {J.}~\bibnamefont {Gu\'ena}},
  \bibinfo {author} {\bibfnamefont {M.}~\bibnamefont {Abgrall}}, \bibinfo
  {author} {\bibfnamefont {S.}~\bibnamefont {Bize}},\ and\ \bibinfo {author}
  {\bibfnamefont {P.}~\bibnamefont {Wolf}},\ }\href
  {https://doi.org/10.1103/PhysRevLett.117.061301} {\bibfield  {journal}
  {\bibinfo  {journal} {Phys. Rev. Lett.}\ }\textbf {\bibinfo {volume} {117}},\
  \bibinfo {pages} {061301} (\bibinfo {year} {2016}{\natexlab{a}})}\BibitemShut
  {NoStop}%
\bibitem [{\citenamefont {Fedderke}\ \emph {et~al.}(2019)\citenamefont
  {Fedderke}, \citenamefont {Graham},\ and\ \citenamefont
  {Rajendran}}]{Fedderke:2019ajk}%
  \BibitemOpen
  \bibfield  {author} {\bibinfo {author} {\bibfnamefont {M.~A.}\ \bibnamefont
  {Fedderke}}, \bibinfo {author} {\bibfnamefont {P.~W.}\ \bibnamefont
  {Graham}},\ and\ \bibinfo {author} {\bibfnamefont {S.}~\bibnamefont
  {Rajendran}},\ }\href {https://doi.org/10.1103/PhysRevD.100.015040}
  {\bibfield  {journal} {\bibinfo  {journal} {Phys. Rev. D}\ }\textbf {\bibinfo
  {volume} {100}},\ \bibinfo {pages} {015040} (\bibinfo {year} {2019})},\
  \Eprint {https://arxiv.org/abs/1903.02666} {arXiv:1903.02666 [astro-ph.CO]}
  \BibitemShut {NoStop}%
\bibitem [{\citenamefont {Ade}\ \emph {et~al.}(2022)\citenamefont {Ade},
  \citenamefont {Ahmed}, \citenamefont {Amiri}, \citenamefont {Barkats},
  \citenamefont {Basu~Thakur}, \citenamefont {Bischoff}, \citenamefont {Beck},
  \citenamefont {Bock}, \citenamefont {Boenish}, \citenamefont {Bullock},
  \citenamefont {Buza}, \citenamefont {Cheshire}, \citenamefont {Connors},
  \citenamefont {Cornelison}, \citenamefont {Crumrine}, \citenamefont
  {Cukierman}, \citenamefont {Denison}, \citenamefont {Dierickx}, \citenamefont
  {Duband}, \citenamefont {Eiben}, \citenamefont {Fatigoni}, \citenamefont
  {Filippini}, \citenamefont {Fliescher}, \citenamefont {Goeckner-Wald},
  \citenamefont {Goldfinger}, \citenamefont {Grayson}, \citenamefont {Grimes},
  \citenamefont {Hall}, \citenamefont {Halal}, \citenamefont {Halpern},
  \citenamefont {Hand}, \citenamefont {Harrison}, \citenamefont {Henderson},
  \citenamefont {Hildebrandt}, \citenamefont {Hilton}, \citenamefont {Hubmayr},
  \citenamefont {Hui}, \citenamefont {Irwin}, \citenamefont {Kang},
  \citenamefont {Karkare}, \citenamefont {Karpel}, \citenamefont {Kefeli},
  \citenamefont {Kernasovskiy}, \citenamefont {Kovac}, \citenamefont {Kuo},
  \citenamefont {Lau}, \citenamefont {Leitch}, \citenamefont {Lennox},
  \citenamefont {Megerian}, \citenamefont {Minutolo}, \citenamefont {Moncelsi},
  \citenamefont {Nakato}, \citenamefont {Namikawa}, \citenamefont {Nguyen},
  \citenamefont {O'Brient}, \citenamefont {Ogburn}, \citenamefont {Palladino},
  \citenamefont {Prouve}, \citenamefont {Pryke}, \citenamefont {Racine},
  \citenamefont {Reintsema}, \citenamefont {Richter}, \citenamefont
  {Schillaci}, \citenamefont {Schwarz}, \citenamefont {Schmitt}, \citenamefont
  {Sheehy}, \citenamefont {Soliman}, \citenamefont {Germaine}, \citenamefont
  {Steinbach}, \citenamefont {Sudiwala}, \citenamefont {Teply}, \citenamefont
  {Thompson}, \citenamefont {Tolan}, \citenamefont {Tucker}, \citenamefont
  {Turner}, \citenamefont {Umilt\`a}, \citenamefont {Verg\`es}, \citenamefont
  {Vieregg}, \citenamefont {Wandui}, \citenamefont {Weber}, \citenamefont
  {Wiebe}, \citenamefont {Willmert}, \citenamefont {Wong}, \citenamefont {Wu},
  \citenamefont {Yang}, \citenamefont {Yoon}, \citenamefont {Young},
  \citenamefont {Yu}, \citenamefont {Zeng}, \citenamefont {Zhang},\ and\
  \citenamefont {Zhang}}]{PhysRevD.105.022006}%
  \BibitemOpen
  \bibfield  {author} {\bibinfo {author} {\bibfnamefont {P.~A.~R.}\
  \bibnamefont {Ade}}, \bibinfo {author} {\bibfnamefont {Z.}~\bibnamefont
  {Ahmed}}, \bibinfo {author} {\bibfnamefont {M.}~\bibnamefont {Amiri}},
  \bibinfo {author} {\bibfnamefont {D.}~\bibnamefont {Barkats}}, \bibinfo
  {author} {\bibfnamefont {R.}~\bibnamefont {Basu~Thakur}}, \bibinfo {author}
  {\bibfnamefont {C.~A.}\ \bibnamefont {Bischoff}}, \bibinfo {author}
  {\bibfnamefont {D.}~\bibnamefont {Beck}}, \bibinfo {author} {\bibfnamefont
  {J.~J.}\ \bibnamefont {Bock}}, \bibinfo {author} {\bibfnamefont
  {H.}~\bibnamefont {Boenish}}, \bibinfo {author} {\bibfnamefont
  {E.}~\bibnamefont {Bullock}}, \bibinfo {author} {\bibfnamefont
  {V.}~\bibnamefont {Buza}}, \bibinfo {author} {\bibfnamefont {J.~R.}\
  \bibnamefont {Cheshire}}, \bibinfo {author} {\bibfnamefont {J.}~\bibnamefont
  {Connors}}, \bibinfo {author} {\bibfnamefont {J.}~\bibnamefont {Cornelison}},
  \bibinfo {author} {\bibfnamefont {M.}~\bibnamefont {Crumrine}}, \bibinfo
  {author} {\bibfnamefont {A.}~\bibnamefont {Cukierman}}, \bibinfo {author}
  {\bibfnamefont {E.~V.}\ \bibnamefont {Denison}}, \bibinfo {author}
  {\bibfnamefont {M.}~\bibnamefont {Dierickx}}, \bibinfo {author}
  {\bibfnamefont {L.}~\bibnamefont {Duband}}, \bibinfo {author} {\bibfnamefont
  {M.}~\bibnamefont {Eiben}}, \bibinfo {author} {\bibfnamefont
  {S.}~\bibnamefont {Fatigoni}}, \bibinfo {author} {\bibfnamefont {J.~P.}\
  \bibnamefont {Filippini}}, \bibinfo {author} {\bibfnamefont {S.}~\bibnamefont
  {Fliescher}}, \bibinfo {author} {\bibfnamefont {N.}~\bibnamefont
  {Goeckner-Wald}}, \bibinfo {author} {\bibfnamefont {D.~C.}\ \bibnamefont
  {Goldfinger}}, \bibinfo {author} {\bibfnamefont {J.}~\bibnamefont {Grayson}},
  \bibinfo {author} {\bibfnamefont {P.}~\bibnamefont {Grimes}}, \bibinfo
  {author} {\bibfnamefont {G.}~\bibnamefont {Hall}}, \bibinfo {author}
  {\bibfnamefont {G.}~\bibnamefont {Halal}}, \bibinfo {author} {\bibfnamefont
  {M.}~\bibnamefont {Halpern}}, \bibinfo {author} {\bibfnamefont
  {E.}~\bibnamefont {Hand}}, \bibinfo {author} {\bibfnamefont {S.}~\bibnamefont
  {Harrison}}, \bibinfo {author} {\bibfnamefont {S.}~\bibnamefont {Henderson}},
  \bibinfo {author} {\bibfnamefont {S.~R.}\ \bibnamefont {Hildebrandt}},
  \bibinfo {author} {\bibfnamefont {G.~C.}\ \bibnamefont {Hilton}}, \bibinfo
  {author} {\bibfnamefont {J.}~\bibnamefont {Hubmayr}}, \bibinfo {author}
  {\bibfnamefont {H.}~\bibnamefont {Hui}}, \bibinfo {author} {\bibfnamefont
  {K.~D.}\ \bibnamefont {Irwin}}, \bibinfo {author} {\bibfnamefont
  {J.}~\bibnamefont {Kang}}, \bibinfo {author} {\bibfnamefont {K.~S.}\
  \bibnamefont {Karkare}}, \bibinfo {author} {\bibfnamefont {E.}~\bibnamefont
  {Karpel}}, \bibinfo {author} {\bibfnamefont {S.}~\bibnamefont {Kefeli}},
  \bibinfo {author} {\bibfnamefont {S.~A.}\ \bibnamefont {Kernasovskiy}},
  \bibinfo {author} {\bibfnamefont {J.~M.}\ \bibnamefont {Kovac}}, \bibinfo
  {author} {\bibfnamefont {C.~L.}\ \bibnamefont {Kuo}}, \bibinfo {author}
  {\bibfnamefont {K.}~\bibnamefont {Lau}}, \bibinfo {author} {\bibfnamefont
  {E.~M.}\ \bibnamefont {Leitch}}, \bibinfo {author} {\bibfnamefont
  {A.}~\bibnamefont {Lennox}}, \bibinfo {author} {\bibfnamefont {K.~G.}\
  \bibnamefont {Megerian}}, \bibinfo {author} {\bibfnamefont {L.}~\bibnamefont
  {Minutolo}}, \bibinfo {author} {\bibfnamefont {L.}~\bibnamefont {Moncelsi}},
  \bibinfo {author} {\bibfnamefont {Y.}~\bibnamefont {Nakato}}, \bibinfo
  {author} {\bibfnamefont {T.}~\bibnamefont {Namikawa}}, \bibinfo {author}
  {\bibfnamefont {H.~T.}\ \bibnamefont {Nguyen}}, \bibinfo {author}
  {\bibfnamefont {R.}~\bibnamefont {O'Brient}}, \bibinfo {author}
  {\bibfnamefont {R.~W.}\ \bibnamefont {Ogburn}}, \bibinfo {author}
  {\bibfnamefont {S.}~\bibnamefont {Palladino}}, \bibinfo {author}
  {\bibfnamefont {T.}~\bibnamefont {Prouve}}, \bibinfo {author} {\bibfnamefont
  {C.}~\bibnamefont {Pryke}}, \bibinfo {author} {\bibfnamefont
  {B.}~\bibnamefont {Racine}}, \bibinfo {author} {\bibfnamefont {C.~D.}\
  \bibnamefont {Reintsema}}, \bibinfo {author} {\bibfnamefont {S.}~\bibnamefont
  {Richter}}, \bibinfo {author} {\bibfnamefont {A.}~\bibnamefont {Schillaci}},
  \bibinfo {author} {\bibfnamefont {R.}~\bibnamefont {Schwarz}}, \bibinfo
  {author} {\bibfnamefont {B.~L.}\ \bibnamefont {Schmitt}}, \bibinfo {author}
  {\bibfnamefont {C.~D.}\ \bibnamefont {Sheehy}}, \bibinfo {author}
  {\bibfnamefont {A.}~\bibnamefont {Soliman}}, \bibinfo {author} {\bibfnamefont
  {T.~S.}\ \bibnamefont {Germaine}}, \bibinfo {author} {\bibfnamefont
  {B.}~\bibnamefont {Steinbach}}, \bibinfo {author} {\bibfnamefont {R.~V.}\
  \bibnamefont {Sudiwala}}, \bibinfo {author} {\bibfnamefont {G.~P.}\
  \bibnamefont {Teply}}, \bibinfo {author} {\bibfnamefont {K.~L.}\ \bibnamefont
  {Thompson}}, \bibinfo {author} {\bibfnamefont {J.~E.}\ \bibnamefont {Tolan}},
  \bibinfo {author} {\bibfnamefont {C.}~\bibnamefont {Tucker}}, \bibinfo
  {author} {\bibfnamefont {A.~D.}\ \bibnamefont {Turner}}, \bibinfo {author}
  {\bibfnamefont {C.}~\bibnamefont {Umilt\`a}}, \bibinfo {author}
  {\bibfnamefont {C.}~\bibnamefont {Verg\`es}}, \bibinfo {author}
  {\bibfnamefont {A.~G.}\ \bibnamefont {Vieregg}}, \bibinfo {author}
  {\bibfnamefont {A.}~\bibnamefont {Wandui}}, \bibinfo {author} {\bibfnamefont
  {A.~C.}\ \bibnamefont {Weber}}, \bibinfo {author} {\bibfnamefont {D.~V.}\
  \bibnamefont {Wiebe}}, \bibinfo {author} {\bibfnamefont {J.}~\bibnamefont
  {Willmert}}, \bibinfo {author} {\bibfnamefont {C.~L.}\ \bibnamefont {Wong}},
  \bibinfo {author} {\bibfnamefont {W.~L.~K.}\ \bibnamefont {Wu}}, \bibinfo
  {author} {\bibfnamefont {H.}~\bibnamefont {Yang}}, \bibinfo {author}
  {\bibfnamefont {K.~W.}\ \bibnamefont {Yoon}}, \bibinfo {author}
  {\bibfnamefont {E.}~\bibnamefont {Young}}, \bibinfo {author} {\bibfnamefont
  {C.}~\bibnamefont {Yu}}, \bibinfo {author} {\bibfnamefont {L.}~\bibnamefont
  {Zeng}}, \bibinfo {author} {\bibfnamefont {C.}~\bibnamefont {Zhang}},\ and\
  \bibinfo {author} {\bibfnamefont {S.}~\bibnamefont {Zhang}} (\bibinfo
  {collaboration} {BICEP/Keck Collaboration}),\ }\href
  {https://doi.org/10.1103/PhysRevD.105.022006} {\bibfield  {journal} {\bibinfo
   {journal} {Phys. Rev. D}\ }\textbf {\bibinfo {volume} {105}},\ \bibinfo
  {pages} {022006} (\bibinfo {year} {2022})}\BibitemShut {NoStop}%
\bibitem [{\citenamefont {Ferguson}\ \emph {et~al.}(2022)\citenamefont
  {Ferguson} \emph {et~al.}}]{SPT-3G:2022ods}%
  \BibitemOpen
  \bibfield  {author} {\bibinfo {author} {\bibfnamefont {K.~R.}\ \bibnamefont
  {Ferguson}} \emph {et~al.} (\bibinfo {collaboration} {SPT-3G}),\ }\href
  {https://doi.org/10.1103/PhysRevD.106.042011} {\bibfield  {journal} {\bibinfo
   {journal} {Phys. Rev. D}\ }\textbf {\bibinfo {volume} {106}},\ \bibinfo
  {pages} {042011} (\bibinfo {year} {2022})},\ \Eprint
  {https://arxiv.org/abs/2203.16567} {arXiv:2203.16567 [astro-ph.CO]}
  \BibitemShut {NoStop}%
\bibitem [{\citenamefont {Ivanov}\ \emph {et~al.}(2019)\citenamefont {Ivanov},
  \citenamefont {Kovalev}, \citenamefont {Lister}, \citenamefont {Panin},
  \citenamefont {Pushkarev}, \citenamefont {Savolainen},\ and\ \citenamefont
  {Troitsky}}]{Ivanov:2018byi}%
  \BibitemOpen
  \bibfield  {author} {\bibinfo {author} {\bibfnamefont {M.~M.}\ \bibnamefont
  {Ivanov}}, \bibinfo {author} {\bibfnamefont {Y.~Y.}\ \bibnamefont {Kovalev}},
  \bibinfo {author} {\bibfnamefont {M.~L.}\ \bibnamefont {Lister}}, \bibinfo
  {author} {\bibfnamefont {A.~G.}\ \bibnamefont {Panin}}, \bibinfo {author}
  {\bibfnamefont {A.~B.}\ \bibnamefont {Pushkarev}}, \bibinfo {author}
  {\bibfnamefont {T.}~\bibnamefont {Savolainen}},\ and\ \bibinfo {author}
  {\bibfnamefont {S.~V.}\ \bibnamefont {Troitsky}},\ }\href
  {https://doi.org/10.1088/1475-7516/2019/02/059} {\bibfield  {journal}
  {\bibinfo  {journal} {JCAP}\ }\textbf {\bibinfo {volume} {02}},\ \bibinfo
  {pages} {059}},\ \Eprint {https://arxiv.org/abs/1811.10997} {arXiv:1811.10997
  [astro-ph.CO]} \BibitemShut {NoStop}%
\bibitem [{\citenamefont {Castillo}\ \emph {et~al.}(2022)\citenamefont
  {Castillo}, \citenamefont {Martin-Camalich}, \citenamefont {Terol-Calvo},
  \citenamefont {Blas}, \citenamefont {Caputo}, \citenamefont {Santos},
  \citenamefont {Sberna}, \citenamefont {Peel},\ and\ \citenamefont {Rubi\~no
  Mart\'\i{}n}}]{Castillo:2022zfl}%
  \BibitemOpen
  \bibfield  {author} {\bibinfo {author} {\bibfnamefont {A.}~\bibnamefont
  {Castillo}}, \bibinfo {author} {\bibfnamefont {J.}~\bibnamefont
  {Martin-Camalich}}, \bibinfo {author} {\bibfnamefont {J.}~\bibnamefont
  {Terol-Calvo}}, \bibinfo {author} {\bibfnamefont {D.}~\bibnamefont {Blas}},
  \bibinfo {author} {\bibfnamefont {A.}~\bibnamefont {Caputo}}, \bibinfo
  {author} {\bibfnamefont {R.~T.~G.}\ \bibnamefont {Santos}}, \bibinfo {author}
  {\bibfnamefont {L.}~\bibnamefont {Sberna}}, \bibinfo {author} {\bibfnamefont
  {M.}~\bibnamefont {Peel}},\ and\ \bibinfo {author} {\bibfnamefont {J.~A.}\
  \bibnamefont {Rubi\~no Mart\'\i{}n}},\ }\href
  {https://doi.org/10.1088/1475-7516/2022/06/014} {\bibfield  {journal}
  {\bibinfo  {journal} {JCAP}\ }\textbf {\bibinfo {volume} {06}}\bibfield
  {number} {\bibinfo  {number} { (06)},\ \bibinfo {pages} {014}},\ }\Eprint
  {https://arxiv.org/abs/2201.03422} {arXiv:2201.03422 [astro-ph.CO]}
  \BibitemShut {NoStop}%
\bibitem [{\citenamefont {Hees}\ \emph
  {et~al.}(2016{\natexlab{b}})\citenamefont {Hees}, \citenamefont {Gu\'ena},
  \citenamefont {Abgrall}, \citenamefont {Bize},\ and\ \citenamefont
  {Wolf}}]{Hees:2016gop}%
  \BibitemOpen
  \bibfield  {author} {\bibinfo {author} {\bibfnamefont {A.}~\bibnamefont
  {Hees}}, \bibinfo {author} {\bibfnamefont {J.}~\bibnamefont {Gu\'ena}},
  \bibinfo {author} {\bibfnamefont {M.}~\bibnamefont {Abgrall}}, \bibinfo
  {author} {\bibfnamefont {S.}~\bibnamefont {Bize}},\ and\ \bibinfo {author}
  {\bibfnamefont {P.}~\bibnamefont {Wolf}},\ }\href
  {https://doi.org/10.1103/PhysRevLett.117.061301} {\bibfield  {journal}
  {\bibinfo  {journal} {Phys. Rev. Lett.}\ }\textbf {\bibinfo {volume} {117}},\
  \bibinfo {pages} {061301} (\bibinfo {year} {2016}{\natexlab{b}})},\ \Eprint
  {https://arxiv.org/abs/1604.08514} {arXiv:1604.08514 [gr-qc]} \BibitemShut
  {NoStop}%
\bibitem [{\citenamefont {{Wilczynska}}\ \emph {et~al.}(2015)\citenamefont
  {{Wilczynska}}, \citenamefont {{Webb}}, \citenamefont {{King}}, \citenamefont
  {{Murphy}}, \citenamefont {{Bainbridge}},\ and\ \citenamefont
  {{Flambaum}}}]{2015MNRAS.454.3082W}%
  \BibitemOpen
  \bibfield  {author} {\bibinfo {author} {\bibfnamefont {M.~R.}\ \bibnamefont
  {{Wilczynska}}}, \bibinfo {author} {\bibfnamefont {J.~K.}\ \bibnamefont
  {{Webb}}}, \bibinfo {author} {\bibfnamefont {J.~A.}\ \bibnamefont {{King}}},
  \bibinfo {author} {\bibfnamefont {M.~T.}\ \bibnamefont {{Murphy}}}, \bibinfo
  {author} {\bibfnamefont {M.~B.}\ \bibnamefont {{Bainbridge}}},\ and\ \bibinfo
  {author} {\bibfnamefont {V.~V.}\ \bibnamefont {{Flambaum}}},\ }\href
  {https://doi.org/10.1093/mnras/stv2148} {\bibfield  {journal} {\bibinfo
  {journal} {MNRAS}\ }\textbf {\bibinfo {volume} {454}},\ \bibinfo {pages}
  {3082} (\bibinfo {year} {2015})},\ \Eprint {https://arxiv.org/abs/1510.02536}
  {arXiv:1510.02536 [astro-ph.CO]} \BibitemShut {NoStop}%
\bibitem [{\citenamefont {Riehle}\ \emph {et~al.}(2018)\citenamefont {Riehle},
  \citenamefont {Gill}, \citenamefont {Arias},\ and\ \citenamefont
  {Robertsson}}]{Riehle_2018}%
  \BibitemOpen
  \bibfield  {author} {\bibinfo {author} {\bibfnamefont {F.}~\bibnamefont
  {Riehle}}, \bibinfo {author} {\bibfnamefont {P.}~\bibnamefont {Gill}},
  \bibinfo {author} {\bibfnamefont {F.}~\bibnamefont {Arias}},\ and\ \bibinfo
  {author} {\bibfnamefont {L.}~\bibnamefont {Robertsson}},\ }\href
  {https://doi.org/10.1088/1681-7575/aaa302} {\bibfield  {journal} {\bibinfo
  {journal} {Metrologia}\ }\textbf {\bibinfo {volume} {55}},\ \bibinfo {pages}
  {188} (\bibinfo {year} {2018})}\BibitemShut {NoStop}%
\bibitem [{\citenamefont {Safronova}\ \emph {et~al.}(2011)\citenamefont
  {Safronova}, \citenamefont {Kozlov},\ and\ \citenamefont
  {Clark}}]{PhysRevLett.107.143006}%
  \BibitemOpen
  \bibfield  {author} {\bibinfo {author} {\bibfnamefont {M.~S.}\ \bibnamefont
  {Safronova}}, \bibinfo {author} {\bibfnamefont {M.~G.}\ \bibnamefont
  {Kozlov}},\ and\ \bibinfo {author} {\bibfnamefont {C.~W.}\ \bibnamefont
  {Clark}},\ }\href {https://doi.org/10.1103/PhysRevLett.107.143006} {\bibfield
   {journal} {\bibinfo  {journal} {Phys. Rev. Lett.}\ }\textbf {\bibinfo
  {volume} {107}},\ \bibinfo {pages} {143006} (\bibinfo {year}
  {2011})}\BibitemShut {NoStop}%
\bibitem [{\citenamefont {Brewer}\ \emph {et~al.}(2019)\citenamefont {Brewer},
  \citenamefont {Chen}, \citenamefont {Hankin}, \citenamefont {Clements},
  \citenamefont {Chou}, \citenamefont {Wineland}, \citenamefont {Hume},\ and\
  \citenamefont {Leibrandt}}]{PhysRevLett.123.033201}%
  \BibitemOpen
  \bibfield  {author} {\bibinfo {author} {\bibfnamefont {S.~M.}\ \bibnamefont
  {Brewer}}, \bibinfo {author} {\bibfnamefont {J.-S.}\ \bibnamefont {Chen}},
  \bibinfo {author} {\bibfnamefont {A.~M.}\ \bibnamefont {Hankin}}, \bibinfo
  {author} {\bibfnamefont {E.~R.}\ \bibnamefont {Clements}}, \bibinfo {author}
  {\bibfnamefont {C.~W.}\ \bibnamefont {Chou}}, \bibinfo {author}
  {\bibfnamefont {D.~J.}\ \bibnamefont {Wineland}}, \bibinfo {author}
  {\bibfnamefont {D.~B.}\ \bibnamefont {Hume}},\ and\ \bibinfo {author}
  {\bibfnamefont {D.~R.}\ \bibnamefont {Leibrandt}},\ }\href
  {https://doi.org/10.1103/PhysRevLett.123.033201} {\bibfield  {journal}
  {\bibinfo  {journal} {Phys. Rev. Lett.}\ }\textbf {\bibinfo {volume} {123}},\
  \bibinfo {pages} {033201} (\bibinfo {year} {2019})}\BibitemShut {NoStop}%
\bibitem [{\citenamefont {{Guo}}\ \emph {et~al.}(2010)\citenamefont {{Guo}},
  \citenamefont {{Wang}},\ and\ \citenamefont {{Ye}}}]{2010JPhB...43m5004G}%
  \BibitemOpen
  \bibfield  {author} {\bibinfo {author} {\bibfnamefont {K.}~\bibnamefont
  {{Guo}}}, \bibinfo {author} {\bibfnamefont {G.}~\bibnamefont {{Wang}}},\ and\
  \bibinfo {author} {\bibfnamefont {A.}~\bibnamefont {{Ye}}},\ }\href
  {https://doi.org/10.1088/0953-4075/43/13/135004} {\bibfield  {journal}
  {\bibinfo  {journal} {Journal of Physics B Atomic Molecular Physics}\
  }\textbf {\bibinfo {volume} {43}},\ \bibinfo {eid} {135004} (\bibinfo {year}
  {2010})}\BibitemShut {NoStop}%
\bibitem [{\citenamefont {{Bothwell}}\ \emph {et~al.}(2019)\citenamefont
  {{Bothwell}}, \citenamefont {{Kedar}}, \citenamefont {{Oelker}},
  \citenamefont {{Robinson}}, \citenamefont {{Bromley}}, \citenamefont {{Tew}},
  \citenamefont {{Ye}},\ and\ \citenamefont {{Kennedy}}}]{2019Metro..56f5004B}%
  \BibitemOpen
  \bibfield  {author} {\bibinfo {author} {\bibfnamefont {T.}~\bibnamefont
  {{Bothwell}}}, \bibinfo {author} {\bibfnamefont {D.}~\bibnamefont {{Kedar}}},
  \bibinfo {author} {\bibfnamefont {E.}~\bibnamefont {{Oelker}}}, \bibinfo
  {author} {\bibfnamefont {J.~M.}\ \bibnamefont {{Robinson}}}, \bibinfo
  {author} {\bibfnamefont {S.~L.}\ \bibnamefont {{Bromley}}}, \bibinfo {author}
  {\bibfnamefont {W.~L.}\ \bibnamefont {{Tew}}}, \bibinfo {author}
  {\bibfnamefont {J.}~\bibnamefont {{Ye}}},\ and\ \bibinfo {author}
  {\bibfnamefont {C.~J.}\ \bibnamefont {{Kennedy}}},\ }\href
  {https://doi.org/10.1088/1681-7575/ab4089} {\bibfield  {journal} {\bibinfo
  {journal} {Metrologia}\ }\textbf {\bibinfo {volume} {56}},\ \bibinfo {eid}
  {065004} (\bibinfo {year} {2019})},\ \Eprint
  {https://arxiv.org/abs/1906.06004} {arXiv:1906.06004 [physics.atom-ph]}
  \BibitemShut {NoStop}%
\bibitem [{\citenamefont {McGrew}\ \emph {et~al.}(2018)\citenamefont {McGrew}
  \emph {et~al.}}]{McGrew:2018mqk}%
  \BibitemOpen
  \bibfield  {author} {\bibinfo {author} {\bibfnamefont {W.~F.}\ \bibnamefont
  {McGrew}} \emph {et~al.},\ }\href {https://doi.org/10.1038/s41586-018-0738-2}
  {\bibfield  {journal} {\bibinfo  {journal} {Nature}\ }\textbf {\bibinfo
  {volume} {564}},\ \bibinfo {pages} {87} (\bibinfo {year} {2018})},\ \Eprint
  {https://arxiv.org/abs/1807.11282} {arXiv:1807.11282 [physics.atom-ph]}
  \BibitemShut {NoStop}%
\bibitem [{\citenamefont {Anastassopoulos}\ \emph {et~al.}(2017)\citenamefont
  {Anastassopoulos} \emph {et~al.}}]{CAST:2017uph}%
  \BibitemOpen
  \bibfield  {author} {\bibinfo {author} {\bibfnamefont {V.}~\bibnamefont
  {Anastassopoulos}} \emph {et~al.} (\bibinfo {collaboration} {CAST}),\ }\href
  {https://doi.org/10.1038/nphys4109} {\bibfield  {journal} {\bibinfo
  {journal} {Nature Phys.}\ }\textbf {\bibinfo {volume} {13}},\ \bibinfo
  {pages} {584} (\bibinfo {year} {2017})},\ \Eprint
  {https://arxiv.org/abs/1705.02290} {arXiv:1705.02290 [hep-ex]} \BibitemShut
  {NoStop}%
\bibitem [{\citenamefont {{Sisk-Reyn{\'e}s}}\ \emph {et~al.}(2022)\citenamefont
  {{Sisk-Reyn{\'e}s}}, \citenamefont {{Matthews}}, \citenamefont {{Reynolds}},
  \citenamefont {{Russell}}, \citenamefont {{Smith}},\ and\ \citenamefont
  {{Marsh}}}]{2022MNRAS.510.1264S}%
  \BibitemOpen
  \bibfield  {author} {\bibinfo {author} {\bibfnamefont {J.}~\bibnamefont
  {{Sisk-Reyn{\'e}s}}}, \bibinfo {author} {\bibfnamefont {J.~H.}\ \bibnamefont
  {{Matthews}}}, \bibinfo {author} {\bibfnamefont {C.~S.}\ \bibnamefont
  {{Reynolds}}}, \bibinfo {author} {\bibfnamefont {H.~R.}\ \bibnamefont
  {{Russell}}}, \bibinfo {author} {\bibfnamefont {R.~N.}\ \bibnamefont
  {{Smith}}},\ and\ \bibinfo {author} {\bibfnamefont {M.~C.~D.}\ \bibnamefont
  {{Marsh}}},\ }\href {https://doi.org/10.1093/mnras/stab3464} {\bibfield
  {journal} {\bibinfo  {journal} {MNRAS}\ }\textbf {\bibinfo {volume} {510}},\
  \bibinfo {pages} {1264} (\bibinfo {year} {2022})},\ \Eprint
  {https://arxiv.org/abs/2109.03261} {arXiv:2109.03261 [astro-ph.HE]}
  \BibitemShut {NoStop}%
\bibitem [{\citenamefont {O'Hare}(2020)}]{ciaran_o_hare_2020_3932430}%
  \BibitemOpen
  \bibfield  {author} {\bibinfo {author} {\bibfnamefont {C.}~\bibnamefont
  {O'Hare}},\ }\href {https://doi.org/10.5281/zenodo.3932430} {\bibinfo {title}
  {cajohare/axionlimits: Axionlimits}} (\bibinfo {year} {2020})\BibitemShut
  {NoStop}%
\end{thebibliography}%

\end{document}